\documentstyle[preprint,eqsecnum,aps]{revtex}
\tightenlines

\input{psfig}

\begin{document}

\title{Conductance of carbon nanotubes with disorder: A numerical study}

\author{ M. P. Anantram\cite{byline1}\\}
\address{ NASA Ames Research Center, Mail Stop T27A-1,
           Moffett Field, CA, USA 94045-1000  }
\author{ T. R. Govindan\cite{byline2}\\}
\address{ Applied Research Laboratory, 
          PO Box 30, State College, PA, USA 16804-0030}

\maketitle

\begin{abstract}
We study the conductance of carbon nanotube wires in the presence of 
disorder, in the limit of phase coherent transport. For this purpose,
we have developed a simple numerical procedure to compute transmission
through carbon nanotubes and related structures. Two models of 
disorder are considered, weak uniform disorder and isolated strong 
scatterers. In the case of weak uniform disorder, our simulations show
that the conductance is not significantly affected by disorder when
the Fermi energy is close to the band center. Further, the transmission 
around the band center depends on the diameter of these zero bandgap 
wires. We also find that the calculated small bias conductance as a 
function of the Fermi energy exhibits a dip when the Fermi energy is
close to the second subband minima. In the presence of strong isolated
disorder, our calculations show a transmission gap at the band center,
and the corresponding conductance is very small. 
\end{abstract}

\vspace{1.0in}

{(\it Article appeared in Physical Review B volume 58, page 4882 (1998))}

\newpage

\section{ Introduction}

The experimental and theoretical study of carbon nanotubes (CNT) has
recently been active because these low dimensional materials display 
interesting properties both from a fundamental physics and applications
view point. The mechanical strength of CNT combined with their rich 
electronic properties have led to demonstrations of their application
as STM tips \cite{Dai96}, field emission sources \cite{Heer95} and 
nanoscale devices \cite{Tans97,Collins97,Chico96a}. CNT can presently 
be cut to lengths varying from tens of a nanometer to a many microns,
and experiments have shown promise as molecular wires \cite{Tans97}. On 
the theoretical side, studies of the conductance of CNT with single 
defects and a junction between tubes have generated interest 
\cite{Chico96b,Saito96,Tamura97}, as has the low energy excitation 
spectrum in the presence of electron-electron interaction 
\cite{Lin97,Egger97,Kane97,Balents97,Krotov97}.

A metallic CNT has two propagating subbands at the Fermi energy. This
can yield a maximum low bias conductance of $\frac{4e^2}{h}$ ($6.25 k
\Omega$). The prospect of realizing conductances close to 
$\frac{4e^2}{h}$ will significantly depend on (i) the role of 
disorder/defects in reducing the conductance of this low dimensional 
material and (ii) ability to realize near perfect contacts with 
macroscopic sized voltage pads. 
Using numerical simulation, we study the effect of
two types of disorder. The first type of disorder is a relatively weak
uniform disorder that is distributed throughout the sample. This model 
has been considered previously in different contexts 
\cite{Beenakker97}. The second type of disorder is isolated strong 
scatterers. These scatterers physically correspond to lattice sites 
onto which an electron cannot hop easily. We find that the two types of
disorder affect the conductance in very different manners. We present 
the results of our conductance calculations in nanotubes of different 
lengths and diameters. We also make suggestions to observe some of 
these results experimentally. The second contribution of our paper is
a procedure that can be used for the numerical computation of the 
transport properties of CNT with defects, T, Y and other junctions
\cite{Zhou95,Menon97,Han98}
and CNT-heterostructures. Our procedure includes the effect of semi 
infinite leads in an efficient manner. The Green's function based 
transport formulation of Refs. \cite{Caroli71,Meir92,Datta_book} is 
employed and is applicable to devices with arbitrary disordered regions 
and junctions.

The paper is organized as follows. We discuss the model and the Green's
function method in section II. This is followed by a discussion of 
the numerical results in section III. We conclude in section IV.

\section{\bf Model}
\label{sect:Model}

The electronic properties of CNT have been calculated in the context
of various approximations. We use the simplest model which assumes the
nanotube to be an $sp^2$ bonded network. The corresponding single 
particle Hamiltonian is \cite{Dresselhaus_book,Charlier96,Louie96},
\begin{eqnarray}
H = \sum_i \epsilon^o_i c_i^\dagger c_i + 
                        \sum_{i,j} t_{ij} c_i^\dagger c_j \mbox{ . }
\label{eq:Hamiltonian}
\end{eqnarray}
Here, $\epsilon^o_i$ is the on-site potential and $t_{ij}$ is the
hopping parameter between lattice sites $i$ and $j$. $\{c_i^\dagger, 
\;c_i\}$ are the creation and annihilation operators at site $i$. In 
the absence of defects, the on-site potential $\epsilon^o_i$ is
zero and the hopping parameter is $-3.1eV$ \cite{Charlier96}. We 
calculate the conductance of a structure that consists of two 
semi-infinite perfect CNT leads separated by a region with defects
(Fig. 1). In the presence of defects, both the on-site potential and
the hopping parameter change. Here, we only consider the variation
in the on-site potential,
\begin{eqnarray}
\epsilon^o_i \rightarrow \epsilon^o_i + \delta \epsilon_i \mbox{.}
\label{eq:model}
\end{eqnarray}
In the case of a uniformly distributed weak disorder, $\delta 
\epsilon_i$ is randomly chosen from the interval $\pm 
|\epsilon_{random}|$ at every lattice point. Increasing 
$\epsilon_{random}$ corresponds to increasing the amount of disorder.
In the case of substitutional defects, $\delta \epsilon_i$ is set to a
large number at some random lattice sites. In a real sample, $\delta
\epsilon_i$ would be expected to have a finite spatial extent. In this
paper, the finite spatial extent is neglected and the random component
is treated as a delta function potential.

The transmission coefficient between the left and right leads is 
calculated using the expression \cite{Meir92,Datta_book},
\begin{eqnarray}
T(E) = trace ( \Gamma_L G^r  \Gamma_R  G^a  ) \mbox{ .} 
\label{eq:Trans}
\end{eqnarray}
The coupling of the device to the left and right leads, $\Gamma_L$ and
$\Gamma_R$ is given by,
\begin{eqnarray}
\Gamma_k (E) = 2 \pi V_k^\dagger Im ( g^r_k (E) ) V_k
\; \mbox{ ,} \label{eq:Gamma}
\end{eqnarray}
where $k \in L,\;R $. $g^r_k (E)$ is the Green's function matrix of the
$k$th semi infinite lead, $G^r$ and $G^a$ are the retarded and advanced
Green's function matrices of the device (including the coupling to the 
semi infinite leads), and $V_k$ is the matrix that couples the 
$k$th lead to device (disordered) region. The trace is over the 
device nodes. To obtain the Green's functions, we solve the 
following equation,
\begin{eqnarray}
\left( EI - H - \Sigma^r_L - \Sigma^r_R \right) G^r = I \mbox{ ,} 
\label{eq:Gr}
\end{eqnarray}
where, $\Sigma^r_k = V_k^\dagger g^r_k V_k$ ($k \in L, R$) represents 
the self 
energy due to the semi infinite leads and $I$ is the identity matrix of
dimension equal to the number of device lattice sites. In general for
a structure with $N$ atoms, solving for all elements of the Green's 
function involves inverting a $N \times N$ matrix. Computational 
resources limit the size of the system that can be considered. 
However, by careful ordering of lattice sites the matrix corresponding
to Eq. (\ref{eq:Gr}) is block tridiagonal [Eq. (\ref{eq:mat})]:
\begin{eqnarray}
\left(
\begin{array}{ccccccc}
A_1    & B_{12}    & O      & O       & O       & O        & O        \\
B_{21} & A_2       & B_{23} & O       & O       & O        & O \\
O      & B_{32}    & \bullet& \bullet & O       & O        & O        \\
O      & O         & \bullet& \bullet & \bullet & O       & O        \\
O      & O         & O      & \bullet & \bullet & \bullet  & O        \\
O      & O         & O      & O       & \bullet & \bullet  & B_{N-1N} \\
O      & O         & O      & O       & O       & B_{N-1N} & A_N
\end{array}
\right)
\left(
\begin{array}{l}
G_{11}^r \\ G_{12}^r \\ \bullet   \\ \bullet   \\ \bullet   \\ G_{1N-1}^r
\\
G_{1N}
\end{array}
\right)
=
\left(
\begin{array}{c}
1 \\ O \\ O \\ O \\ O \\ O \\ O
\end{array}
\right) \mbox{ .}        \label{eq:mat}
\end{eqnarray}
For this purpose we divide the structure into smaller units, each unit
typically representing one/few rings  of atoms along the circumference
of the tube.  The diagonal submatrix $A_i$ (dimension of $N_i \times 
N_i$) represents $EI - H - \Sigma^r_L - \Sigma^r_R$ of the $i$th unit
and the off diagonal submatrix $B_{ij}$ (dimension of $N_i \times N_j$)
represents the coupling between units $i$ and $j$, where $N_i$ and
$N_j$ are the number of sites in units $i$ and $j$. $O$ are empty
matrices. In the near neighbor tight binding scheme, $B_{ij}$ is
non zero only when $|i-j|=1$. Hence, a block tridiagonal structure for
Eq. (\ref{eq:Gr}). Calculating the phase coherent transmission 
coefficient involves only the off diagonal component of the Green's
function connecting the left and right ends of the device ($G^r_{N1}$).
This further reduces the labor to compute the transmission coefficient.
We solve for $G^r_{N1}$ by using an efficient block tridiagonal
elimination procedure. Using this procedure, we are able to calculate
the transmission coefficient through long disordered regions.

The Green's function $g^r_k$ is calculated via an iterative procedure
\cite{Samanta98}. The matrix equation corresponding to the semi 
infinite leads is the same as Eq. (\ref{eq:mat}), only that the matrix 
is semi infinite, with all $A_i = A = E - H + i \eta$ (evaluated at a 
unit in lead $k$) and $B_{ij} = B_{ji}^t = B$. The equations for 
$\Gamma_k$ and $\Sigma_k^r$ involve only the submatrix $[g_k^r]_{11}$, 
which corresponds to the semi infinite Green's function of the unit
in lead $k$ that is closest to the device region. From Eq. 
(\ref{eq:mat}), $[g_{k}^r]_{11}$ is given by the following 
equation \cite{Samanta98},
\begin{eqnarray}
[g_k^r]_{11} = \frac{I}{E - H + i\eta - B^t [g_k^r]_{11} B}  
				\mbox{  .} \label{eq:g}
\end{eqnarray}

The current across the device is calculated using the Landauer-Buttiker
formula,
\begin{eqnarray}
I = \frac{2e}{\hbar} \int dE \; T(E) [ f_1(E) - f_2(E) ] \mbox{ ,}
\label{eq:current}
\end{eqnarray}
where, the factor $2$ accounts for spin degeneracy. $f_1(E)$ and 
$f_2(E)$ are the Fermi functions of the waves incident from the two 
contacts to the device. Note that in the present work, we calculate 
only the phase coherent transmission coefficient (the effect of 
electron-phonon interaction is neglected) and that temperature 
dependence is only via the Fermi factors of electrons.
Two important considerations in a calculation of current are the
equilibrium location of the Fermi level with respect to the band 
bottom of the device when connected to the contacts \cite{Datta97} 
and the self-consistent potential profile of the device in the presence 
of an applied bias. We assume the case of reflectionless contacts 
\cite{Datta_book,Tekman91} and consider the scenario where the Fermi 
energy can be varied with respect to the band bottom of the CNT. 
The ability to vary the Fermi energy in a CNT has been demonstrated
experimentally in Refs. \cite{Tans97,Bockrath97,Bezryadin98}.
The potential in the device is not calculated self-consistently and we
simply assume a linear drop in the applied potential, while calculating
the current versus voltage characteristics.

\section{\bf Results and Discussion}

\subsection{\bf Weak uniform  disorder}

In a conventional one dimensional chain, electrons traverse only a 
single effective path across the leads and as a result transmission
is significantly altered by small amounts of disorder 
\cite{Ziman_book}. In comparison, electrons in a CNT can travel around 
defects because of the larger number of atoms in a cross section (the 
number of modes is only two at the band center). An important issue is
how disorder affects the conductance of CNT wires. We calculate 
transmission (by this we mean the sum of the transmission coefficient
over the incident modes, $\sum_n T_n$) as a function of both the length
of the disordered region and the magnitude of disorder using the 
procedure described in section \ref{sect:Model}. Transmission 
versus energy and conductance versus gate voltage for one configuration
of disorder is shown in Fig. 2. Transmission in a CNT has the following
features that are in common with a single moded one dimensional chain:
rapidly varying peaks that signify local resonances created by disorder
and decrease in the average value with increasing disorder as the 
mismatch in the energies of the resonances increases with increase in
disorder \cite{Beenakker97,Ziman_book}.

We now discuss features that are typical of carbon nanotubes.
Fig. 2 shows a significant reduction in the
transmission coefficient at energies close to the beginning of the 
second subband, even for weak disorder strengths. This leads to a dip
in conductance when the Fermi level is close to the beginning of
the second subband (Fig. 3). The origin of this dip is due to low 
velocity electrons in the second subband and can be understood as 
follows. In a perfect lattice, the velocity ($dE/dk$) of electrons with
the quantum number of the second subband and with an energy close to 
the beginning of the second subband is nearly zero. These low velocity 
electrons are easily reflected by the smallest of disorders. 
Disorder causes mixing of the first and second subbands. As a result,
electrons incident in either subband at these energies develop a large 
reflection coefficient (in comparison to energies close to the 
band center). Increasing the disorder strength results in further 
reduction of the conductance and also results in the broadening of the 
dip. Subsequent to Eq. (\ref{eq:model}), we mentioned that the finite
spatial extent of $\delta \epsilon_i$ is neglected in our study. 
A model that includes the finite spatial extent of $\delta \epsilon_i$
would require larger lengths of disordered regions to
see dips whose magnitude is comparable to those in Figs. 2 and 3.
The results in Fig. 2 are for one random 
configuration of disorder distributed over a length of 1000 $\AA$. We
have carried out simulations over different length scales and disorder 
configurations and our results for the average transmission at the band
center, averaged over more than a thousand disorder configurations are
summarized in Fig. 4. The important point here is that for the smaller
disorder strengths, the average transmission of a micron long (10,10)
tube is not significantly affected by disorder, thus demonstrating the
relative robustness of transport at the band center. For disordered
regions larger than some localization length ($L_0$), the conductance
of quasi one dimensional samples has been predicted to decrease 
exponentially with length, $g=g_0\;exp(-L/L_0)$, in the phase coherent
limit \cite{Beenakker97}. For lengths
shorter than the localization length, the decrease in conductance is
not given by this equation. We observe this to be the case in our
simulations (inset of Fig. 4). The value of $L_0$ corresponding to 
disorder strengths of 1eV and 1.75eV are 3353 $\AA$ and 1383 $\AA$ 
respectively.

We also compute transmission for nanotubes of
different diameters. This study illustrates the effect of the number
of atoms in a cross section of the wire. We compare transmission 
of the (10,10) tube with that of a (5,5) and (12,0) zig zag tubes.
The diameters of these tubes are 13.4 $\AA$, 9.4 $\AA$ and 6.7 $\AA$
respectively.  For the (10,10) and (5,5) tubes, the band structure at 
energies close to the Fermi energy are similar
\cite{Dresselhaus_book}. But the number of atoms in a unit cell of a
(5,5) tube is only half of that in a (10,10) tube (they have 20 and 
40 atoms respectively). Fig. 5 shows the average transmission versus 
wire length. The important point here is that in spite of the identical
transmission of a disorder free (10,10) and (5,5) tube at energies 
around the band center, transmission is smaller for the (5,5) tube 
in the presence of disorder. This is because the (5,5) tube has a 
smaller number of atoms around the circumference, thus reducing the 
number of paths by which electrons can travel around defects and across
the device. To support this view point, we compare these results 
to conductance of a 1000 $\AA$ long (12,0) zig zag tube. 
We find that transmission is in between that of the (10,10) and (5,5) 
tubes (Fig. 5). This is because the (12,0) tube has a diameter 
that is in between 
that of the (10,10) and (5,5) tube, and as a result the number of 
effective paths is larger than that available to a (5,5) tube but 
smaller than that of a (10,10) tube.

Can any of these effects be observed experimentally? Recently, arm 
chair, zig zag and tubes with chiralities in between have been 
experimentally characterized by STM imaging \cite{Wildoer98,Odom98}.
Transport 
measurements of single wall CNT at low temperatures has so far been
limited by coulomb blockade due to large barriers at the contact-CNT 
interface \cite{Tans97,Bockrath97}. Disorder of some degree is bound to
exist in CNT samples and we believe that the variation in the linear 
response conductance with the gate potential\cite{footnote1} and the dip
in the conductance at energies close to the crossing of the first and 
second subbands can be observed in situations where the contact
resistance is not the dominant factor. The length dependence of the 
conductance can also be studied by varying the length of the tube 
between the electrodes. One caveat is that phonon scattering will cause
an increase in the low bias conductance in the presence of strong 
disorder with an increase in temperature. Our calculations are
relevant at low temperatures where phonon scattering is not 
significant. 

\subsection{\bf Strong isolated defects}

An electron cannot hop on to such a defect site either due to a large
mismatch in the on site potential or weak bonds with its neighbors
(section II). Scattering from a single defect causes a
maximum reduction in the transmission at the band center E=0. For
example, the transmission of a (10,10) tube reduces from 2 to 
approximately 0.94 due to a single defect \cite{Chico96a}. We are 
interested in the effect of a few such defects scattered randomly along
the length of the tube. Reflection from more than a single defect 
causes the creation of quasi bound states along the tube, the exact 
locations of which are sensitive to the position of the defects. We find
that a significant feature that is independent of the exact location of 
these defects is the opening of a {\it transmission gap} at the center
of the band as defects are added. The second feature that we see in the 
simulations is that the width of the transmission gap increases with 
increase in the defect density. The transmission has sharp decreases at
energies corresponding to the opening of the second subband, but this 
effect is relatively weak compared to the previous case of disorder.
The simulation results illustrating these features are  shown in Fig. 
6 for a wire of length 1000 $\AA$ with ten defects scattered along 
the length randomly. As a result of the transmission gap, the low bias
conductance is greatly reduced from the defect free case, at zero gate
voltage.
Conductance further depends significantly on temperature (inset of Fig.
7a).  In summary, while the conductance is not significantly affected
by relatively weak uniform disorder (Figs. 3 and 4), we find that the
conductance here is much smaller than $2e^2/h$ at zero gate voltage. 
Conductance increases with gate voltage, with features of 
resonances due to the quasi bound states superimposed. These features 
get averaged out with increase in temperature. We also calculate 
current as a function of applied voltage by assuming a linear drop in 
the applied voltage.
Transmission  at each applied voltage is computed and then we use
Eq. (\ref{eq:current}) to calculate the current. The main feature in
the I-V characteristic is the small increase in current with applied 
voltage close to the zero of applied voltage (Fig. 8). The experimental
work in Ref. \cite{Collins97} measured the I-V characteristics of a 
CNT rope. One of their main findings was that the differential 
conductance is very small at zero bias and that it increases with an 
increase in applied bias. The qualitative features of Fig. 8 are 
similar but an important difference is that the experiments were
performed on a rope of single walled tubes, in which case it has
recently been predicted that a band gap could open due to tube-tube
interactions \cite{Delaney98}.

\section{\bf Conclusions}

We present a method to calculate the phase coherent 
transmission through nanotubes using a Green's function formalism that
can include the effect of semi infinite leads and can handle many 
defects and junctions with relative ease. We use this formalism 
to study the importance of scattering due to disorder. Two simple 
models of disorder are considered and their effect on the conductance
discussed. In the presence of weak uniform
disorder, we find that the conductance is not significantly affected by
disorder and that the wires behave as reasonably good quantum wires.
For example a micron long (10,10) CNT with a disorder strength of 1eV
(section II) has a conductance comparable to $0.16 \frac{e^2}{h}$. We
predict that an experiment involving measurement of conductance
versus gate voltage will show a dip in conductance when the Fermi
energy is close to the opening of the second subband (Fig. 3).
We compare the conductance of wires with varying diameters and find
that the transmission (conductance) increases with the diameter of the
tube for a given disorder strength (Fig. 5; note that in the absence
of disorder the conductance is independent of the tube diameter at 
zero gate voltage). We attribute this to a decrease in the number of 
effective paths by which an electron can traverse across the device 
with decrease in the diameter.
The second type of defect considered is strong isolated scatterers. In
contrast to the previous type of disorder, this disorder creates a gap
in the transmission at the band center and a corresponding large
reduction in the low bias conductance. Such disorder would destroy the
good conductance properties of the wire at the band center. The work
presented is based on numerical simulations. Of
interest could be further conductance experiments to look for features
described in this paper.
Carbon nanotubes provide an unprecedented natural scenario for wires 
with a few modes and a relatively small cross sectional area. An 
analytical study of the effect of disorder in these
systems and the dependence of the conductance as a function of diameter 
and chirality would be useful. Also, of interest for future work would 
be a study that includes the effect of phonon scattering. 

\section{\bf Acknowledgments}
We would like to thank Jie Han (NASA Ames Research Center) for sharing
his expertise on many aspects of carbon nanotubes and for lively 
discussions.
We would like to thank Manoj Samanta and Supriyo Datta of Purdue
University for communicating the result of (Eq. \ref{eq:g}) before
publication \cite{Samanta98}. It is also a pleasure to acknowledge
useful discussions with Supriyo Datta (Purdue University) and 
Mathieu Kemp (Northwestern University). 
\nopagebreak


\pagebreak

\noindent
{\bf Figure Captions:}

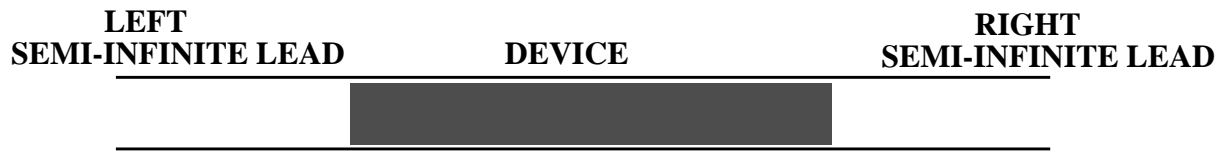
Fig. 1: A schematic representation of the structure across which the
transmission is calculated. Our calculation account for semi infinite
leads connected to the disordered region.

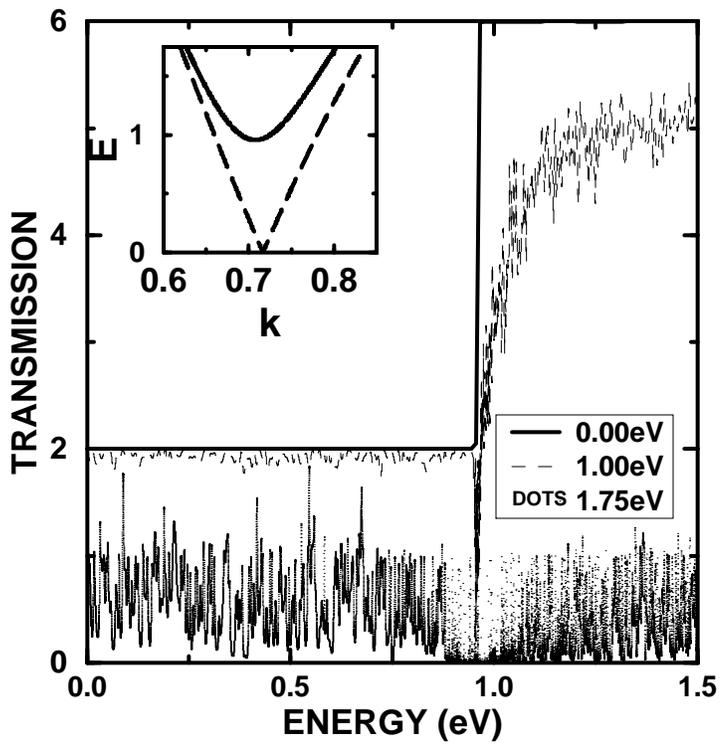
Fig. 2: Transmission versus energy for a (10,10) CNT with disorder
distributed over a length of 1000 $\AA$.
The significant features here are the robustness of the transmission
around the zero of energy, as the strength of disorder is increased
and the dip in transmission at energies close to the beginning of the 
second subband. The inset shows energy versus wave vector for the first 
(solid) and the second subband (dashed); the velocity of electrons at
the minima of the solid line is zero.

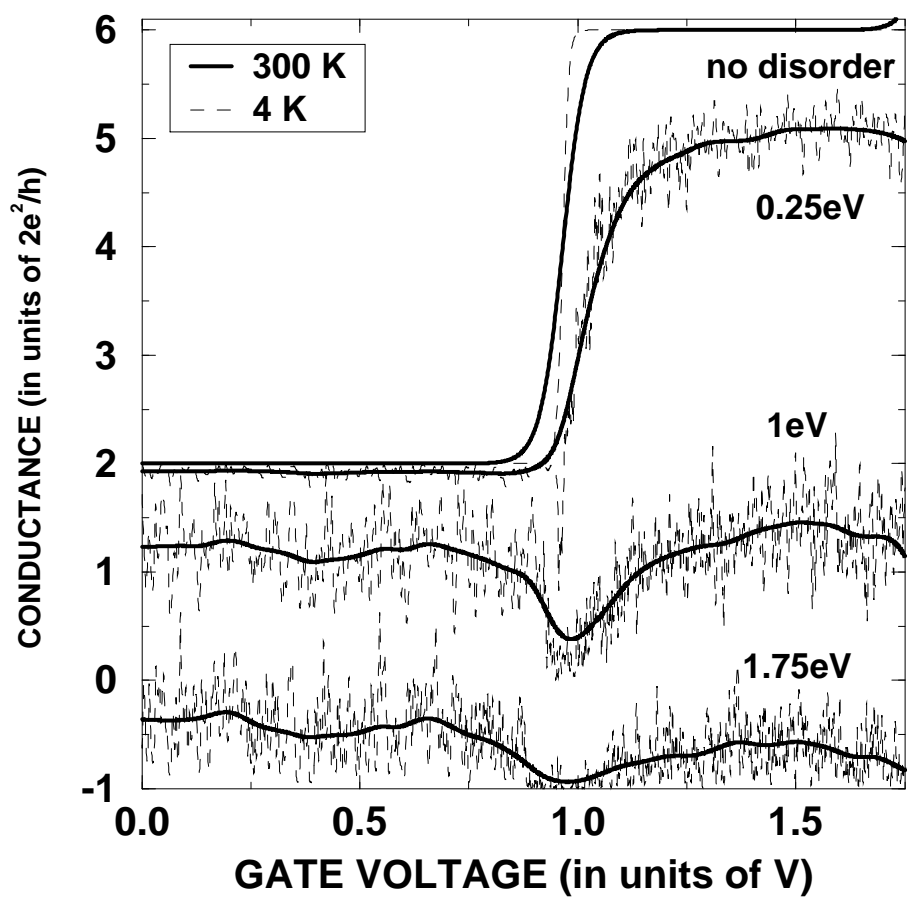
Fig. 3: The low bias conductance versus gate voltage for the structure
used in Fig. 2. The figure clearly shows the dip in the conductance
when the fermi energy is close to the second subband minima. At the
lower temperature, features due to the quasi bound resonances in the
disordered region are not averaged out when compared to the high
temperature case.

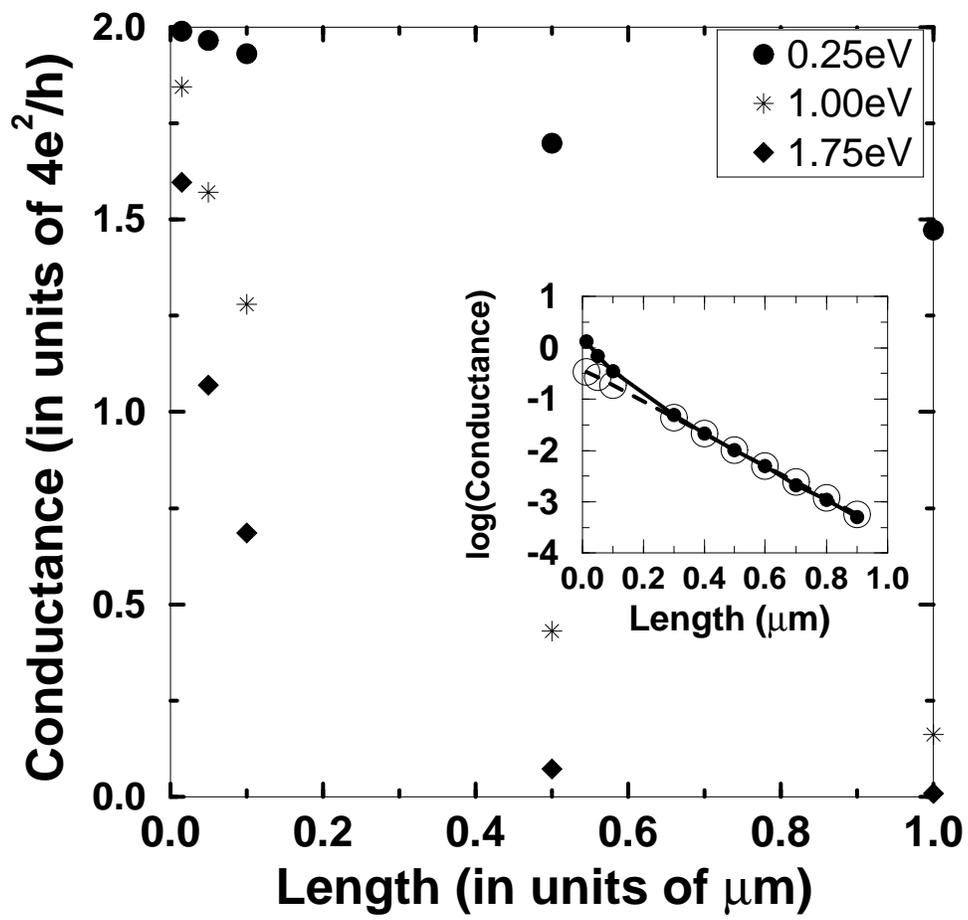
Fig. 4: The conductance versus length of the (10,10) CNT. While for
the large disorder strengths, the conductance is significantly
affected by disorder, the conductance is reasonably large for the
smaller values of disorder. This demonstrates the robustness of
these wires to weak uniform disorder. Inset: log(Conductance) versus
length for disorder strength of 1.75eV in a (5,5) CNT. The solid 
line/filled circle
corresponds to the simulation and the dashed line/empty circle
corresponds to that obtained using $g=g_0 exp(-L/L_0)$. 

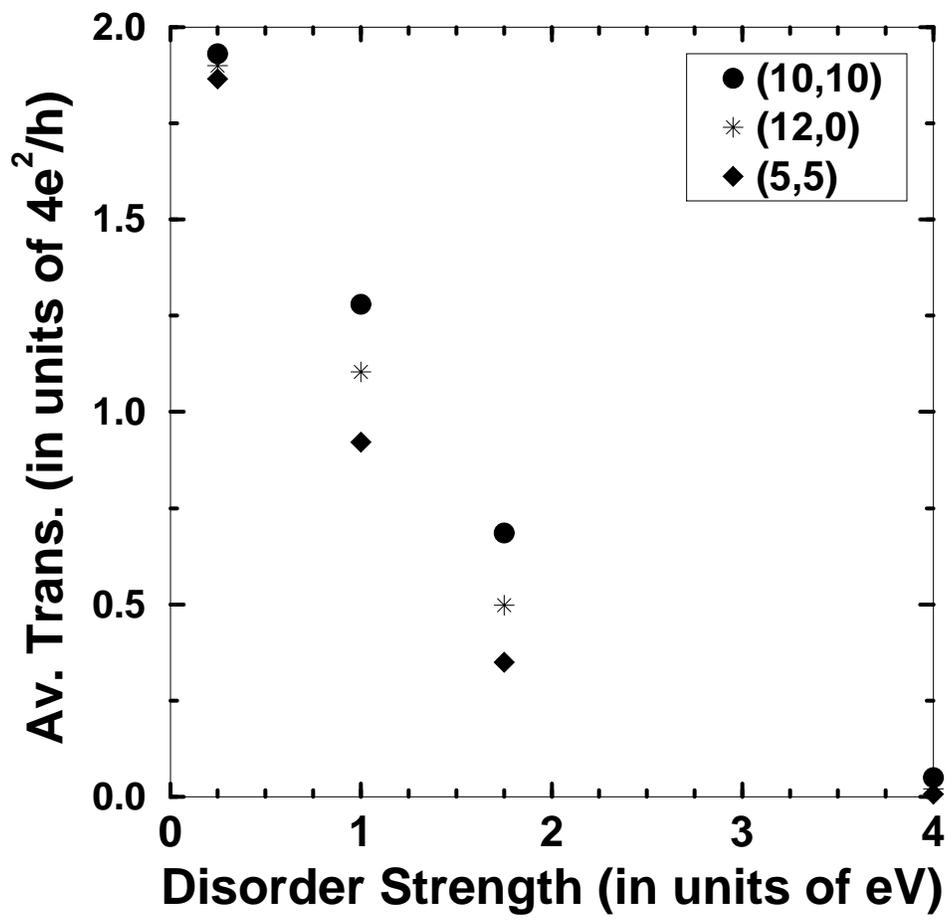
Fig. 5: The average transmission at the band center versus disorder 
strength for wires of different diameter and chirality; the 
transmission has been averaged over a thousand different realizations
of the disorder. The main feature here is that the average
transmission decreases with a decrease in the number of atoms along the 
circumference of the wire (see text).

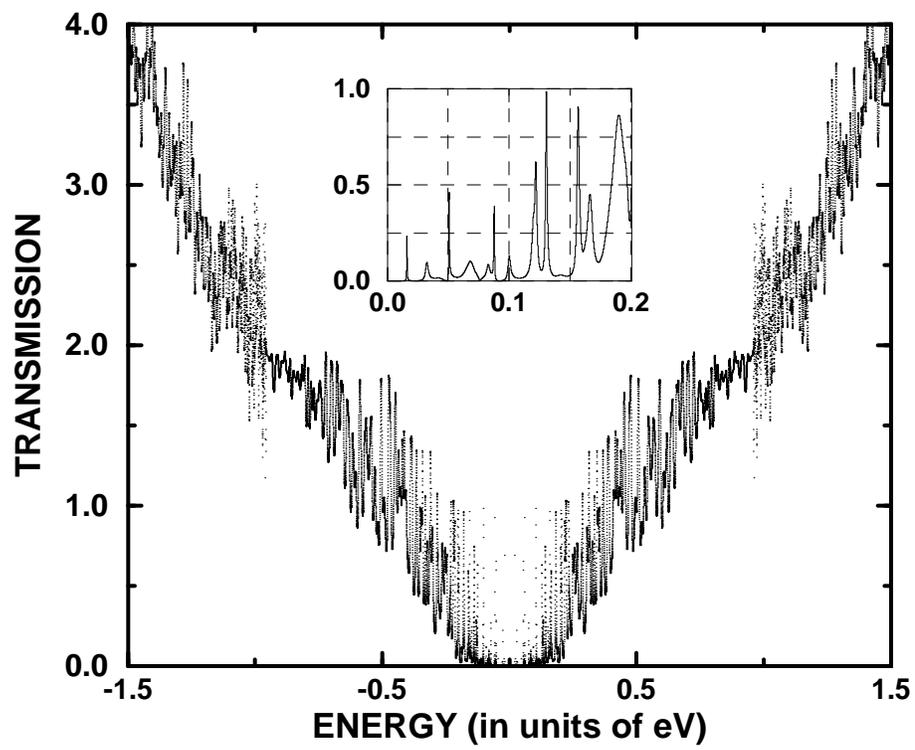
Fig. 6: The transmission versus energy for a (10,10) CNT with ten
strong isolated scatterers sprinkled randomly along a length of 1000
$\AA$. The main prediction here is the opening of a transmission gap 
around the zero of energy. Inset: Comparison of the transmission for
tubes of lengths 1000 $\AA$ (solid) and 140 $\AA$ (dashed) with ten 
scatterers in each case. The transmission gap is larger for the larger
defect density and the sharp resonances close to the zero of energy 
are suppressed with increasing defect density. 

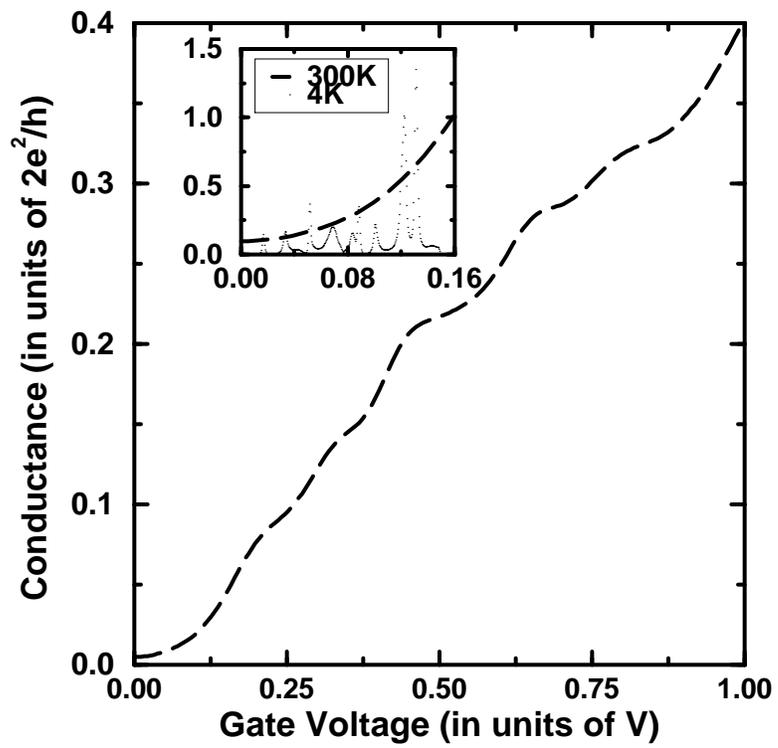
Fig. 7: The conductance at T=300K for the case in Fig. 6. The low
conductance at zero gate bias represents the transmission gap in Fig.
6. The transmission resonances of Fig. 6 get averaged out here. The 
inset compares the effect of temperature on the conductance. Close to 
zero gate voltage, the conductance is clearly suppressed at the lower
temperature.

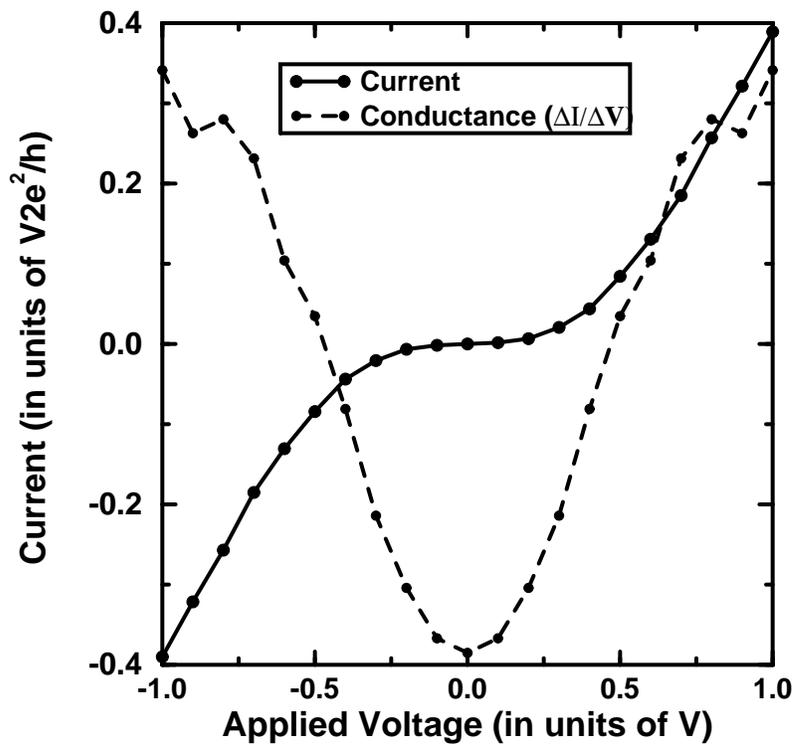
Fig. 8: The current (shifted by $-0.4$ units along the current axis) 
versus applied voltage for the same structure as in Fig. 6. The dashed
curve is the differential conductance, which is very small at low
applied voltages.


\begin{thebibliography}{100}
\bibitem{footnote1} The gate potential causes the fermi energy of the 
CNT to shift with respect to its band bottom.
\end{thebibliography}
\end{document}